\documentclass[aps,prd,reprint,groupedaddress,amsmath,amssymb]{revtex4-2}
\usepackage[absolute]{textpos}

\begin{document}

\begin{textblock*}{3in}(2.75in,0.3in)
PHYSICAL REVIEW D \textbf{107}, L081301 (2023)
\end{textblock*}
\begin{textblock*}{2.5in}(5.55in,10.5in)
\copyright~2023 American Physical Society
\end{textblock*}

\title{Ultraslow PSR J0901-4046 with an ultrahigh magnetic field of $3.2\times10^{16}$~G}
\author{D. N. Sob'yanin}
\email{sobyanin@lpi.ru}
\affiliation{P. N. Lebedev Physical Institute of the Russian Academy of Sciences, Moscow 119991, Russia}

\begin{abstract}
The recent discovery of a radio-emitting neutron star with an ultralong spin period of 76~s, PSR J0901-4046, raises a fundamental question on how such a slowly rotating magnetized object can be active in the radio band. A canonical magnetic field of $1.3\times10^{14}$~G estimated from the pulsar period and its time derivative is wholly insufficient for PSR J0901-4046 to operate. Consideration of a magnetic inclination angle of $10^\circ$ estimated from the pulse width gives a higher magnetic field of $1.5\times10^{15}$~G, which is still an order of magnitude lower than the necessary minimum of $2.5\times10^{16}$~G following from the death line for radio pulsars with magnetic fields exceeding the critical value $4.4\times10^{13}$~G. We show that if the observed microstructure of single pulses reflects relativistic beaming, the inferred surface magnetic field appears to be $3.2\times10^{16}$~G, and without this assumption it is no less than $2.7\times10^{16}$~G, which explains the existence of radio emission from PSR J0901-4046. This estimation makes PSR J0901-4046 a radio pulsar with the strongest magnetic field known and is a sign that PSR J0901-4046 slows down not by magnetic-dipole radiation, but rather by an electric current of 56~MA, when rotational energy is expended in accelerating charged particles over the polar cap.
\end{abstract}

\maketitle

PSR J0901-4046, with a period of $P\approx75.9$~s and a period derivative of $\dot{P}\approx2.25\times10^{-13}\text{ s}\,\text{s}^{-1}$, was discovered on September 27, 2020 at 1284~MHz with the MeerKAT radio telescope and was associated with an ultra-slowly rotating magnetized neutron star \citep{CalebEtal2022}. The canonical estimate $1.3\times10^{14}$~G for the surface magnetic field allows one to classify PSR J0901-4046 as a high-magnetic-field radio pulsar with the field exceeding the critical value $B_\text{cr}=m_e^2c^3/e\hbar\approx4.4\times10^{13}$~G. Since the estimated upper limit on the 0.5--10~keV x-ray luminosity $L_\text{X}\lesssim3.2\times10^{30}\text{ erg}\,\text{s}^{-1}$ exceeds the rotational energy loss $\dot{E}\approx2.0\times10^{28}\text{ erg}\,\text{s}^{-1}$, it is not impossible that PSR J0901-4046 is a magnetar, and its single pulses exhibit a quasiperiodicity, in some sense resembling quasiperiodic oscillations from magnetars \citep{BretzEtal2021,CastroTiradoEtal2021,LiEtal2022}. PSR J0901-4046 rotates more than 3 times slower than the former record holder PSR J0250+5854 with $P=23.5$~s \citep{TanEtal2018,AgarEtal2021} and according to its rotational characteristics should reside in the graveyard of neutron stars in the $P{-}\dot{P}$ or $P{-}B$ parameter space, and thus not emit any radio waves \citep{ChenRuderman1993,IstominSobyanin2007,MorozovaEtal2012,ZhouEtal2017}. In this paper we address the problem of the origin of radio emission from the apparently dead PSR J0901-4046. Eschewing the model of magnetic-dipole radiation and solely using energy transformation during plasma multiplication above the polar cap of a strongly magnetized rotating neutron star, we show that the actual surface magnetic field of PSR J0901-4046 is 2 orders of magnitude higher than the conventional estimate. This removes PSR J0901-4046 from the graveyard and allows it to be active in the radio band.

The observed S-shaped rotational phase dependence of the polarization position angle constrains the impact parameter $\beta\lesssim0.2^\circ$ in the framework of the rotating-vector model \citep{RadhakrishnanCooke1969,Komesaroff1970}, implying a close passage of the line of sight near the magnetic axis, but the magnetic inclination angle $\alpha$, which is highly desirable for an accurate estimation of the magnetic field of the neutron star even when the standard model of an inclined magnetic dipole rotating in the vacuum is considered, remains unconstrained \citep{CalebEtal2022}. To estimate $\alpha$, we use the data on the pulsar period and pulse width.

\citet{Rankin1990} studied the period dependence of core-component half-power widths for radio pulsars with interpulses (which are orthogonal rotators with $\alpha\approx90^\circ$) and on this basis proposed an expression for the half-power width of the core component at 1~GHz in the pulse profile of a generally inclined pulsar,
\begin{equation}
\label{W50Rankin}
W_{50}^\text{1 GHz}=\frac{2.45^\circ P^{-0.5}}{\sin\alpha}.
\end{equation}
For PSR J0901-4046 we probably observe a core-single profile ($\mathrm{S_t}$) in the classification of \citet{Rankin1983}. A typical feature of such profiles is their transformation to triple profiles ($\mathrm{T}$) with increasing observational frequency. In the profiles observed in the L and UHF (ultra-high-frequency) bands (see the Supplementary Data Fig.~2 in Ref.~\citep{CalebEtal2022}) we may notice a hint of small bumps flanking the pulse center, each at a distance of $\sim0.15$~s, which might be a sign of the beginning of the transformation $\mathrm{S_t\rightarrow T}$. Whether this is so can become clear after conducting radio observations of PSR J0901-4046 at higher frequencies.

Fortunately, we need not determine the exact type of the observed profile to be entitled to use Eq.~\eqref{W50Rankin} because the same lower boundary line $W_{50}=2.45^\circ P^{-0.5}$, corresponding to $\alpha=90^\circ$, was found in pulsars with core and conal components at 1~GHz \cite{MaciesiakEtal2012}, and thus $W_{50}$ can be considered the half-power width of the whole profile irrespective of its type. In addition, \citet{MitraEtal2016} through the study of pulsar profiles at 333 and 618~MHz found the lower boundary line $W_{50}=2.7^\circ P^{-0.5}$, which is dominated by 618-MHz measurements and after mapping to 1~GHz using the frequency dependence $W\propto\nu^{-0.19}$ observed by the authors corresponds to the aforementioned boundary $W_{50}=2.45^\circ P^{-0.5}$.

PSR J0901-4046 has the same pulse width $W_{50}\approx0.3\text{ s}\approx1.4^\circ$ in the UHF and L bands without signs of radius-to-frequency mapping, so Eq.~\eqref{W50Rankin} implies $\alpha\approx11^\circ$. The half-power point lies just about the trailing feature in the average profile \cite{CalebEtal2022}, so if we interpret this feature as an emerging side peak of the triple profile, then using $W_{50}$ might not be absolutely reliable because this width in such a case does not contain either the sole core component or the whole profile with core and conal components. To overcome this likely difficulty, we use the tenth-power width $W_{10}$ instead, containing all profile components. \citet{MitraEtal2016} studied the period dependence of $W_{10}$ and found the lower boundary line $W_{10}=5.7^\circ P^{-0.5}$. Mapping it from 618~MHz to 1~GHz using the aforementioned dependence $W\propto\nu^{-0.19}$, we get
\begin{equation}
\label{sinAlphaViaW10}
{\sin\alpha}=\frac{5.2^\circ P^{-0.5}}{W_{10}^\text{1 GHz}}.
\end{equation}
Since $W_{10}\approx0.7\text{ s}\approx3.3^\circ$ for PSR J0901-4046 (see the Supplementary Data Fig.~2 in Ref.~\citep{CalebEtal2022}), the angle between the rotational and magnetic axes is estimated as
\begin{equation}
\label{alphaViaW10}
\alpha\approx10^\circ.
\end{equation}
This value, which will be adopted below, appears to be virtually the same as that from $W_{50}$.

In individual pulses of PSR J0901-4046 we observe a sort of quasiperiodicity with a characteristic period of $P_\mu\sim76$~ms (full range 29--183~ms), and the whole pulse is a set of consecutive distinct micropulses with a median width of $w_\mu\sim49$~ms (full range 13--120~ms) (see the Supplementary Data Fig.~6 in Ref.~\citep{CalebEtal2022}). The observed relation $P_\mu\sim2w_\mu$ is typical for the microstructure of ordinary pulsars \citep{PopovEtal1987}, and the observed values of $w_\mu$ and $P_\mu$ are consistent with the empirical period dependences found earlier for normal and millisecond pulsars: $w_\mu=10^{-3}P\sim76$~ms \citep{Cordes1979}, $w_\mu=(6\pm1)\times10^{-4}P^{1.1\pm0.2}\sim70$~ms \citep{KramerEtal2002}, $P_\mu=10^{-3}(1.1 P + 0.08)\sim84$~ms \citep{MitraEtal2015}, and $P_\mu=(1.06\pm0.62)\times10^{-6}(P/1\text{ ms})^{0.96\pm0.09}\sim51$~ms \citep{DeEtal2016}.

Usually the pulse as a whole consists of subpulses, often attributed to plasma beamlets generated by a rotating carousel of sparks in the inner gap above the polar cap of the neutron star \citep{RudermanSutherland1975}, and the subpulses in turn consist of micropulses. In our case we can estimate the subpulse width through the rotation period as $w_\text{s}=7\times10^{-3}P\sim0.53$~s \citep{Cordes1979}, which appears on the order of the pulse width, thus implying that the whole pulse might be a subpulse. This is not unusual in the framework of a recent single-spark model developed for PSR J2144-3933 \citep{MitraEtal2020}, and the existence of only one spark in the gap of PSR J0901-4046 can be explained analogously, namely, by an extremely narrow polar cap due to very slow rotation, the radius of which is $R_\text{pc}=R_\text{ns}^{1.5}/R_\text{lc}^{0.5}\approx17$~m, where $R_\text{ns}=10$~km is the canonical neutron star radius and $R_\text{lc}=cP/2\pi\approx3.6\times10^{6}$~km is the light cylinder radius, and is even less than $R_\text{pc}\approx50$~m for PSR J2144-3933.

\citet{Gil1982,Gil1986} assumed that the observed width of micropulses is determined by relativistic beaming of curvature radiation from charged particles moving along curved magnetic field lines with almost the speed of light. From this relation it is possible to estimate the Lorentz factors of relativistic particles for pulsars with microstructure, which appear to lie in the range 610--23000 \citep{LangeEtal1998,PopovEtal2002}. Here we use the same idea to determine the Lorentz factor $\gamma$ of secondary particles for PSR J0901-4046, which is necessary for finding its magnetic field. The opening angle of radiation is $\phi_\mu=2/\gamma$ \citep{Bordovitsyn1999} and the observed micropulse width is $w_\mu\text{ (rad)}=\phi_\mu/\sin\alpha$ for $\beta\ll\alpha$; hence,
\begin{equation}
\label{gammaFromWmu}
\gamma=\frac P{\pi w_\mu \sin\alpha}\approx2700.
\end{equation}
This value is comparable to the aforementioned values found earlier for ordinary pulsars.

Meanwhile, there are alternative explanations of microstructure in pulsars \citep{MofizEtal1985,CordesEtal1990,StrohmayerEtal1992,MachabeliEtal2001,AsseoPorzio2006}, so it is desirable to estimate the Lorentz factor in an independent way without relying on a model of the origin of microstructure for some comparison. Before doing so, we need to recall some necessary facts from the theory of plasma generation in ultrahigh magnetic fields (see Ref.~\citep{IstominSobyanin2007} for details). The Lorentz factor \eqref{gammaFromWmu} characterizes the energy of secondary electrons and positrons created as a result of one-photon absorption of high-energy photons in a magnetic field and their transformation to electron-positron pairs, and the photons are in turn generated by primary particles in the acceleration gap over the polar cap through the curvature mechanism. The energy of the particles of the created pair is related to the angle $\chi$ between the directions of the photon wave vector and magnetic field line at the point of photon absorption as $\gamma=1/\chi$. After a curvature photon has been emitted by a primary particle, it travels in the direction tangential to the magnetic field line at the point of emission and $\chi=l/\rho_0(1+l/R_\text{ns})$ first grows linearly with traveling length $l$ while $l\ll R_\text{ns}$, but then approaches a maximum of $\chi_\text{max}=R_\text{ns}/\rho_0$, where $\rho_0=(4/3)(R_\text{ns}R)^{0.5}$ is the radius of curvature of the magnetic field line at the point of emission and $R\geqslant R_\text{lc}$ is the distance at which this dipole line crosses the magnetic equatorial plane.

If the point of emission is at a distance of $s R_\text{pc}$ from the magnetic axis, where $0\leqslant s\leqslant1$, then $R=s^{-2}R_\text{lc}$ and $\rho_0=(4/3s)(R_\text{ns}R_\text{lc})^{0.5}$. \citet{MitraRankin2002} showed from an analysis of observations that $s\approx0.5$ is most appropriate when estimating pulsar emission heights through the conal beam radius, and we adopt this value of~$s$. Therefore, the minimum Lorentz factor $\gamma_\text{min}=1/\chi_\text{max}$ of secondary particles is
\begin{equation}
\label{gammaMin}
\gamma_\text{min}=\frac4{3s}\biggl(\frac{R_\text{lc}}{R_\text{ns}}\biggr)^{0.5}\approx1600.
\end{equation}
The Lorentz factor \eqref{gammaFromWmu} obtained from microstructure characteristics is larger than but close to the value \eqref{gammaMin}, which is independent of assumptions about the mechanism of microstructure formation and is based on purely geometric considerations. Thus, the estimate \eqref{gammaFromWmu} seems realistic, the more so that in view of the extreme rotational characteristics of PSR J0901-4046 its operation is unusual in itself and the Lorentz factor of secondary particles cannot significantly exceed the minimum possible Lorentz factor~\eqref{gammaMin}.

Now we can estimate the energy of photons producing secondary electrons and positrons with the Lorentz factors we just found. We deal with the case of a strong magnetic field $B\gg B_\text{cr}$, where $B_\text{cr}=m_e^2c^3/e\hbar\approx4.4\times10^{13}$~G is the critical magnetic field, and then a photon with energy $\varepsilon_\text{ph}$ emitted by a primary particle accelerated in the gap is absorbed in the magnetic field just after traveling the threshold distance $l_\text{t}=2\rho_0 m_e c^2/\varepsilon_\text{ph}$, i.e., upon approaching the threshold angle $\chi_\text{t}=2 m_e c^2/\varepsilon_\text{ph}$, which means that $\varepsilon_\text{ph}=2\gamma m_e c^2$ and the energy of the photon converts completely to the energy of the produced particles, which do not emit synchrotron photons, contrary to the case of normal pulsars with $B\ll B_\text{cr}$ \citep{IstominSobyanin2007,Timokhin2010,TimokhinHarding2015}. The characteristic energy of curvature radiation from primary particles moving with Lorentz factor $\gamma_0$ along magnetic field lines with radius of curvature $\rho_0$ is $\varepsilon_\text{ph}=(3/2)\hbar c\gamma_0^3/\rho_0$, so the Lorentz factor of primary particles becomes
\begin{equation}
\label{gamma0}
\gamma_0=\biggl(\frac43 \frac{\gamma\rho_0}{\lambdabar}\biggr)^{1/3}\approx5.3\times10^7,
\end{equation}
where $\lambdabar =\hbar/m_e c$ is the Compton wavelength and Eq.~\eqref{gammaFromWmu} has been used. The lower boundary for the Lorentz factor of primary particles not using microstructure characteristics can be found by substituting Eq.~\eqref{gammaMin} into Eq.~\eqref{gamma0}:
\begin{equation}
\label{gamma0min}
\gamma_{0\,\text{min}}=\frac43 s^{-2/3} \biggl( \frac{R_\text{lc}}{\lambdabar}\biggr)^{1/3}\approx4.5\times10^7.
\end{equation}
Naturally, $\gamma_0\gtrsim\gamma_{0\,\text{min}}$.

Now we are in a position to estimate the magnetic field at the stellar surface by relating $\gamma_0$ to the absolute value of the accelerating electric potential $U(s)=U(1-s^2)(1-i_0)\cos\alpha$ in the inner gap of the neutron star \citep{IstominSobyanin2007}, where $i_0=\rho/\rho_\text{GJ}$ is the ratio of the charge density $\rho=e(n_+ - n_-)$ to the Goldreich-Julian density $\rho_\text{GJ}=-B\cos\alpha/cP$, $n_+$ and $n_-$ are the number densities for positrons and electrons, and
\begin{equation}
\label{U}
U=\frac{B R_\text{ns}^3}{2R_\text{lc}^2}
\end{equation}
is the standard potential at the magnetic pole \citep{RudermanSutherland1975}. Note that this potential is not limited by the magnetic field strength and directly follows from solving the Poisson equation. Starting from zero at the stellar surface, the potential gradually increases and finally forms at heights $h$ larger than the characteristic transverse distance, which in our case is the polar cap radius, $h>R_\text{pc}$. To be applicable, the potential \eqref{U} should not be screened by secondary pairs before these heights are reached, which gives the corresponding inequality for the mean free path $l_\text{t}=\rho_0\chi_\text{t}=\rho_0/\gamma$ of photons emitted by primary particles, $l_\text{t}>R_\text{pc}$. We thus get $\gamma<\rho_0/R_\text{pc}=(4/3s)R_\text{lc}/R_\text{ns}\approx9.7\times10^5$, which is satisfied [see Eq.~\eqref{gammaFromWmu}]. Since the sole $l_\text{t}$ appears to exceed $R_\text{pc}$, we need not additionally consider the particle acceleration and radiation formation lengths in the spirit of Ref.~\citep{IstominSobyanin2011} because these contributions can only increase~$h$. We see that PSR J0901-4046 differs from the usually modeled pulsars with thin polar gaps, for which $h\ll R_\text{pc}$ \citep{TimokhinHarding2015,TimokhinHarding2019}.

Since $|n_+-n_-|\leqslant n_+ +n_-$, we have $|i_0|\leqslant i=j/j_\text{GJ}$, where $i$ is the ratio of the absolute value of the real current density $j$ to the formal density $j_\text{GJ}=B\cos\alpha/P$. We will verify below that $i\ll1$ for PSR J0901-4046, so $i_0$ may be dropped and the radial distribution of Lorentz factors over the polar cap is then
\begin{equation}
\label{gamma0s}
\gamma_0(s)=\frac{eU}{m_e c^2}(1-s^2)\cos\alpha.
\end{equation}
Combining Eqs. \eqref{U} and \eqref{gamma0s} and putting $s=0.5$, we finally obtain the surface magnetic field at the magnetic pole of PSR J0901-4046,
\begin{equation}
\label{B}
B=\frac83 \frac{\gamma_0 m_e c^2 R_\text{lc}^2}{e R_\text{ns}^3 \cos\alpha}\approx3.2\times10^{16}\text{ G},
\end{equation}
where we have used the value \eqref{gamma0}.

The inferred magnetic field $B=3.2\times10^{16}$~G [Eq.~\eqref{B}] appears to be 2 orders of magnitude higher than the field $B_\text{st}=3.2\times10^{19}(P\dot{P})^{0.5}\approx1.3\times10^{14}$~G given by the standard magnetic-dipole estimate. Such fields are not prohibited, and there is a hint of large internal magnetic fields of $10^{16}$~G in the magnetars 4U 0142+61, 1E 1547.0-5408, and SGR 1900+14 \citep{MakishimaEtal2014,MakishimaEtal2021,MakishimaEtal2021b}, though the standard estimate only gives fields on the order of $B_\text{st}$ for PSR J0901-4046. An activity of neutron stars with such high surface magnetic fields exceeding $B_\text{cr}$ in the radio band was once considered impossible due to photon splitting \citep{BaringHarding1998}. This effect consists in breaking a photon into two photons with lower energies. As already mentioned, a photon with energy $\varepsilon_\text{ph}$ produces an electron-positron pair upon approaching a threshold angle $\chi_\text{t}\propto\varepsilon_\text{ph}^{-1}$, and photon splitting leads to a decrease in $\varepsilon_\text{ph}$ and a corresponding increase in $\chi_\text{t}$. When $\chi_\text{t}$ becomes larger than $\chi_\text{max}$, secondary pairs are not produced and plasma multiplication together with the concomitant radio emission is impossible.

Against this background the discovery of radio-emitting magnetars was then quite surprising \citep{MalofeevEtal2005,CamiloEtal2006,CamiloEtal2007}. \citet{IstominSobyanin2007} theoretically studied plasma generation in a magnetar magnetosphere taking into account the photon splitting kinetics and explained the radio emission by showing that photon splitting undoubtedly occurs but does not suppress plasma generation because of the specifics of the splitting of photons with different polarizations. The death line in the $P{-}B$ diagram for neutron stars with high magnetic fields $B\gtrsim B_\text{cr}$ has the form [we use Eq.~(64) from Ref.~\citep{IstominSobyanin2007} and adopt $R_\text{ns}=10$~km and $i\ll1$]
\begin{equation}
\label{deathLine}
B\gtrsim\frac{P^{7/3}}{\cos\alpha}10^{12}\text{ G}.
\end{equation}

For PSR J0901-4046, the inequality \eqref{deathLine} gives the condition
\begin{equation}
\label{Bdeath}
B\gtrsim B_\text{death}\approx2.5\times10^{16}\text{ G}.
\end{equation}
The standard estimate $B_\text{st}=1.3\times10^{14}$~G is inconsistent with the operation of PSR J0901-4046. With extra consideration of the inclination angle \eqref{alphaViaW10} and a factor of 2 when passing from the equatorial to polar magnetic field, the maximum that the rotating-magnetic-dipole model can give is $B_\text{d}=2 B_\text{st}/\sin\alpha\approx1.5\times10^{15}$~G, which is still an order of magnitude lower than $B_\text{death}$ and is insufficiently high to explain the observed activity in the radio band. Meanwhile, the surface magnetic field $B=3.2\times10^{16}$~G [Eq.~\eqref{B}] derived above from microstructure characteristics of PSR J0901-4046 satisfies the inequality~\eqref{Bdeath}, which explains the existence of plasma multiplication and the observed radio emission.

The discrepancy between $B_\text{d}$ and $B$ is a sign that PSR J0901-4046 does not slow down due to the magnetic-dipole radiation. An extremely slowly rotating PSR J0901-4046 has a very narrow polar cap of radius $R_\text{pc}\approx17$~m and correspondingly a very large nonvacuum closed magnetosphere, in which the plasma accumulates and which screens the magnetic-dipole radiation. How can we explain the pulsar slow down if the standard picture of an inclined magnetic moment rotating in the vacuum is not valid for PSR J0901-4046? The total power released in an electrodynamic system is generally voltage times electric current, an idea that was applied to other systems \citep{Sobyanin2017,IstominChernyshovSobyanin2020}. Let us split the polar cap into nested rings centered at the magnetic pole and consider a ring of radius $r$ and thickness $dr$. The voltage corresponding to this ring is $V=U[1-(r/R_\text{pc})^2]\cos\alpha$, while the electric current is $dI=2Ir dr /R_\text{pc}^2$, where a uniform current distribution over the polar cap is assumed. The power corresponding to the ring is $dW=V dI=2UI\cos\alpha(1-s^2)s ds$, where $s=r/R_\text{pc}$, so the power corresponding to the whole polar cap is $W=\int dW=(1/2)UI\cos\alpha$. Since the neutron star has two polar caps, the total energy loss becomes
\begin{equation}
\label{EdotUI}
\dot{E}=UI\cos\alpha,
\end{equation}
where $U$ is given by Eq.~\eqref{U}. This relation is consistent with a viewpoint that the pulsar slow down is caused by the current \citep{BeskinGurevichIstomin1984}. Since the observed $\dot{E}\approx2.0\times10^{28}\text{ erg}\,\text{s}^{-1}$, the electric current flowing through the polar cap is
\begin{equation}
\label{I}
I=\frac34 \frac{e\dot{E}}{\gamma_0 m_e c^2}\approx56\text{ MA},
\end{equation}
so that the normalized current is $i=I/I_\text{GJ}\approx4.6\times10^{-5}\ll1$, where
\begin{equation}
\label{Igj}
I_\text{GJ}=j_\text{GJ} \pi R_\text{pc}^2=\frac43 \frac{\gamma_0 m_e c^3}{e}\approx1.2\text{ TA}
\end{equation}
is the formal Goldreich-Julian current. In this model of pulsar slow down the rotational energy of the neutron star fully transforms into the energy of primary particles in the acceleration gap, and not into the energy of magnetic-dipole radiation.

Why does the standard magnetic-dipole estimate not hold if we only consider the dipole magnetic field? The physical reason for pulsar slow down is Lorentz forces acting on a rotating neutron star, not the form of the magnetic field. The pulsar will not slow down if there is no flowing electric current \cite{BeskinGurevichIstomin1983}, so if one has a pulsar with a dipole or a quasidipole magnetic field but no currents, any slow down will be absent. Such a picture does not contradict pulsar slow down in the vacuum, where magnetospheric currents are absent but nonzero Lorentz forces physically giving the common magnetic-dipole slow down appear because of surface charges and currents resulting from field discontinuities at the stellar surface. For our case of a nonvacuum plasma-filled magnetosphere with low dimensionless current $i\ll1$, the pulsar necessarily slows down with a much lower speed than in the vacuum \cite{BeskinGurevichIstomin1983}, which is the reason why formally applying the standard $P{-}\dot{P}$ estimate to PSR J0901-4046 underestimates the surface magnetic field. The standard estimate would only be adequate for large currents $i\sim1$, when the energy loss \eqref{EdotUI} has virtually the same form $\propto B^2\Omega^4 R_\text{ns}^6$ as in the simple magnetic-dipole model.

We see that PSR J0901-4046 lies near the death line, which seems natural in the sense that such high magnetic fields should be very rare. Since we have an independent lower boundary for the Lorentz factor of primary particles, by substituting the value \eqref{gamma0min} into Eq.~\eqref{B} we can also give an estimate from below for the surface magnetic field that is not based on microstructure characteristics of the pulse,
\begin{equation}
\label{Bmin}
B_\text{min}\approx2.7\times10^{16}\text{ G}.
\end{equation}
This lower limit almost coincides with the death value \eqref{Bdeath}.

We have so far assumed canonically that PSR J0901-4046 has a dipole magnetic field, but the possibility of nondipole fields is also discussed in some aspects of the physics of neutron stars \citep{BarsukovTsygan2010,IgoshevEtal2016,KalapotharakosEtal2021}, so it is interesting to see what changes if the pulsar magnetic field is a multipole. Here we restrict ourselves to considering the situation of an axisymmetric quadrupole magnetic field of the form $B_r=3D(3\cos^2\theta-1)/4r^4$ and $B_\theta=3D\sin\theta\cos\theta/2r^4$, where $r$ is the radial coordinate, $\theta$ is the polar angle with respect to the axis of symmetry, and $D$ is the quadrupole moment [see Fig.~1(b) in Ref.~\citep{LongEtal2007}]. We can find that the form of the quadrupole magnetic field lines is given by the equation $r^2=C\sin^2\theta\cos\theta$, where the constant $C$ enumerates the lines and for the last closed magnetic field line, touching the light cylinder, is $C_\text{lc}=(25\sqrt{5}/16)R_\text{lc}^2$. If the point of emission of curvature radiation is at a distance of $s R_{\text{pc}\,\text{q}}$ from the axis, where $R_{\text{pc}\,\text{q}}=(4/5^{1.25})R_\text{ns}^2/R_\text{lc}$ is the quadrupole polar cap radius, the radius of curvature for $s=0.5$ becomes $\rho_{0\,\text{q}}=(5^{1.25}/8s)R_\text{lc}\approx1.9R_\text{lc}$, which together with Eq.~\eqref{gamma0} gives a Lorentz factor of primary particles in the quadrupole case that is an order of magnitude higher than in the dipole case, $\gamma_{0\,\text{q}}\approx4.0\times10^8$. Replacing $U$ in Eq.~\eqref{gamma0s} by the quadrupole potential $U_\text{q}=B_\text{q} R_{\text{pc}\,\text{q}}^2/2R_\text{lc}$, we arrive at an unrealistically strong magnetic field $B_\text{q}\approx3.1\times10^{23}$~G, which indicates that the global magnetic quadrupole structure is impossible in PSR J0901-4046.

The reason why we have obtained such a high quadrupole magnetic field is a smaller curvature of field lines in the polar cap region and, more importantly, a tiny polar cap with radius $R_{\text{pc}\,\text{q}}\approx1.5$~cm incapable of providing the necessary acceleration for reasonable magnetic fields. However, the relativistic conducting plasma flows can deform the fields and open some of the closed magnetic field lines near the boundary between the closed and open magnetospheres, leading to an increase in the accelerating potential. Let us estimate the geometric parameters of the quadrupole that could give the same magnetic field as the dipole. Such a quadrupole should lie in a cylinder with radius $R_\text{q}=\alpha R_\text{lc}$, where $\alpha<1$. Since the radius of curvature of magnetic field lines and the polar cap radius are modified according to the equations $\rho^*_{0\,\text{q}}=\alpha\rho_{0\,\text{q}}$ and $R^*_{\text{pc}\,\text{q}}=\alpha^{-1}R_{\text{pc}\,\text{q}}$, we have $\gamma^*_0=\alpha^{1/3}\gamma_0$ for the Lorentz factor of primary particles and $B^*_\text{q}=\alpha^{7/3}B_\text{q}$ for the magnetic field in the polar region of the quadrupole magnetosphere. From the condition $B^*_\text{q}=B$ we get $\alpha=(B/B_\text{q})^{3/7}\approx10^{-3}$, so the resulting quadrupole may exist up to distances of only $R_\text{q}\approx10^{-3}R_\text{lc}\approx370R_\text{ns}$. The Lorentz factor of primary electrons and positrons $\gamma^*_{0\,\text{q}}\approx4.0\times10^7$ and the polar cap radius $R^*_{\text{pc}}\approx14$~m appear to be similar to those obtained above for the pure dipole.

Beyond the boundary $r\sim R_\text{q}$ the quadrupole magnetic field loses significance and we may expect the existence of some dipole-like magnetic field. Assuming that at this boundary the dipole and quadrupole fields are comparable, with the first $B^*_\text{d}(r)\sim B^*_\text{d}(r/R_\text{ns})^{-3}$ prevailing in the area $r>R_\text{q}$ and the second $B^*_\text{q}(r)\sim B(r/R_\text{ns})^{-4}$ prevailing in the area $r<R_\text{q}$, we can qualitatively estimate an effective magnetic field at the stellar surface that corresponds to the dipole part, $B^*_\text{d}\sim B R_\text{ns}/R_\text{q}\sim8.7\times10^{13}$~G. Interestingly, this field appears to be comparable to the magnetic field $B_\text{st}$ implied by the model of magnetic-dipole loss in the vacuum. Thus, if we speculate beyond the classical dipole fields, it seems plausible that the standard $P{-}\dot{P}$ estimate characterizes the far magnetospheric fields while the near fields are high and multipole.

In conclusion, we have considered a problem concerning the very possibility of radio emission from PSR J0901-4046 with unique rotational characteristics. Its ultraslow rotation implies that the surface magnetic field should exceed the death value $B_\text{death}=2.5\times10^{16}$~G, which is needed for an efficient cascade multiplication of an electron-positron plasma generating radio emission. The standard estimate $B_\text{st}=3.2\times10^{19}(P\dot{P})^{0.5}=1.3\times10^{14}$~G excludes this possibility, but it is not well grounded. We have considered energy transformation during plasma multiplication and established a link between the surface magnetic field and the energies of primary and secondary electrons and positrons. Individual pulses from PSR J0901-4046 are characterized by microstructure, and the model of relativistic beaming relates the micropulse width to the Lorentz factors of secondary particles, which appear to be $\gamma=2700$. These are determined by the energy of curvature radiation going from primary particles accelerated over the polar cap with Lorentz factors of $\gamma_0=5.3\times10^7$. Their energy is determined by the electric potential in the acceleration gap proportional to the surface magnetic field at the magnetic pole, and the latter field appears to be $B=3.2\times10^{16}$~G. Even if we do not use the relativistic beaming model, we cannot arrive at magnetic fields lower than $B_\text{min}=2.7\times10^{16}$~G. This fact explains the existence of radio emission from PSR J0901-4046 and makes it the most magnetized radio pulsar known.

\end{document}